\documentclass[final, journal, twoside]{IEEEtran}

\usepackage{cite}
\usepackage{commath}
\usepackage{amsmath}
\usepackage{amssymb}
\usepackage{amsfonts}
\usepackage{graphicx}
\usepackage{mwe,tikz}
\usepackage[percent]{overpic}
\usepackage{psfrag}
\usepackage{array,multirow,graphicx}
\usepackage{pgfplots}
\usepackage{subfigure}
\newcommand{\emax}{e_{\mathrm{max}}}
\newcommand{\dmax}{d_{\mathrm{max}}}
\newcommand{\rmax}{r_{\mathrm{max}}}

\pgfplotsset{compat=1.14}
\usepackage{scalerel}
\usepackage{tikz}
\usetikzlibrary{svg.path}

\definecolor{orcidlogocol}{HTML}{A6CE39}
\tikzset{
  orcidlogo/.pic={
    \fill[orcidlogocol] svg{M256,128c0,70.7-57.3,128-128,128C57.3,256,0,198.7,0,128C0,57.3,57.3,0,128,0C198.7,0,256,57.3,256,128z};
    \fill[white] svg{M86.3,186.2H70.9V79.1h15.4v48.4V186.2z}
                 svg{M108.9,79.1h41.6c39.6,0,57,28.3,57,53.6c0,27.5-21.5,53.6-56.8,53.6h-41.8V79.1z M124.3,172.4h24.5c34.9,0,42.9-26.5,42.9-39.7c0-21.5-13.7-39.7-43.7-39.7h-23.7V172.4z}
                 svg{M88.7,56.8c0,5.5-4.5,10.1-10.1,10.1c-5.6,0-10.1-4.6-10.1-10.1c0-5.6,4.5-10.1,10.1-10.1C84.2,46.7,88.7,51.3,88.7,56.8z};
  }
}

\newcommand\orcidicon[1]{\href{https://orcid.org/#1}{\mbox{\scalerel*{
\begin{tikzpicture}[yscale=-1,transform shape]
\pic{orcidlogo};
\end{tikzpicture}
}{|}}}}

\usepackage{hyperref} 

\begin{document}

\title{Global Minimax Approximations and Bounds for the Gaussian $Q$-Function by Sums of Exponentials}
\author{Islam~M.~Tanash \orcidicon{0000-0002-9824-6951} and Taneli~Riihonen \orcidicon{0000-0001-5416-5263},~\IEEEmembership{Member,~IEEE}%
\thanks{Manuscript received  June 20, 2019; revised November 8, 2019, March 6, 2020, and May 26, 2020; accepted June 22, 2020. 
This work was supported by the Academy of Finland under the grant 310991/326448. The associate editor coordinating the review of this paper and approving it for publication was F.~J.~L{\'o}pez-Mart{\'i}nez.  {\em (Corresponding author: Islam~M.~Tanash.)}}%
\thanks{The authors are with Unit of Electrical Engineering, Faculty of Information Technology and Communication Sciences, Tampere University, FI-33720 Tampere, Finland (e-mail: islam.tanash@tuni.fi; taneli.riihonen@tuni.fi).}%
\thanks{Digital Object Identifier X}%
}

\markboth{IEEE TRANSACTIONS ON COMMUNICATIONS }%
{Tanash \MakeLowercase{\textit{et al.}}: Global Minimax Approximations and Bounds for the Gaussian $Q$-Function by Sums of Exponentials}

\maketitle

\begin{abstract} 
This paper presents a novel systematic methodology to obtain new simple and tight approximations, lower bounds, and upper bounds for the Gaussian \mbox{$Q$-function}, and functions thereof, in the form of a weighted sum of exponential functions. They are based on minimizing the maximum absolute or relative error, resulting in globally uniform error functions with equalized extrema. In particular, we construct sets of equations that describe the behaviour of the targeted error functions and solve them numerically in order to find the optimized sets of coefficients for the sum of exponentials. This also allows for establishing a trade-off between absolute and relative error by controlling weights assigned to the error functions' extrema. We further extend the proposed procedure to derive approximations and bounds for any polynomial of the \mbox{$Q$-function}, which in turn allows approximating and bounding many functions of the \mbox{$Q$-function} that meet the Taylor series conditions, and consider the integer powers of the \mbox{$Q$-function} as a special case. In the numerical results, other known approximations of the same and different forms as well as those obtained directly from quadrature rules are compared with the proposed approximations and bounds to demonstrate that they achieve increasingly better accuracy in terms of the global error, thus requiring significantly lower number of sum terms to achieve the same level of accuracy than any reference approach of the same form.
\end{abstract}


\section{Introduction}
\label{sec:Introduction}

\IEEEPARstart{T}{he} Gaussian \mbox{$Q$-function} and the related error function $\operatorname{erf}(\cdot)$ are ubiquitous in and fundamental to communication theory, not to mention all other fields of statistical sciences where the Gaussian/normal distribution is often encountered. In particular, the \mbox{$Q$-function} measures the tail probability of a standard normal random variable $X$ having unit variance and zero mean, i.e., $Q(x)=\operatorname{Prob}(X \geq x)$, by which
\begin{subequations}
\label{eq:1}
\begin{align}
Q(x) 
&\triangleq \frac{1}{\sqrt{2\pi}} \int_{x}^{\infty} \exp\left(-{\textstyle \frac{1}{2}} t^2\right)\,\mathrm{d}t\\
&= \frac{1}{\pi} \int_{0}^{\frac{\pi}{2}} \exp\left(-{\textstyle \frac{1}{2\sin^2 \theta}} x^2\right)\,\mathrm{d}\theta
\:\:\text{[for $x \geq 0$]}.\:
\label{eq:2}
\end{align}
\end{subequations}
The latter integral is the so-called Craig's formula\cite{258319,545646546}, obtained by manipulating the original results of\cite{Weinstein1974Sep,78576587}. 

The Gaussian \mbox{$Q$-function} has many applications in statistical performance analysis such as evaluating bit, symbol, and block error probabilities for various digital modulation schemes and different fading models\cite{digibook,79269,1096558,4299416,8425781,809706,752121}, and evaluating the performance of energy detectors for cognitive radio applications \cite{5403561,6068200}, whenever noise and interference or a channel can be modelled as a Gaussian random variable. However, in many cases formulating such probabilities will result in complicated integrals of the \mbox{$Q$-function} that cannot be expressed in a closed form in terms of elementary functions.
Therefore, finding tractable approximations and bounds for the \mbox{$Q$-function} becomes a necessity in order to facilitate expression manipulations and enable its application over a wider range of analytical studies. Toward this demand, several approximations and bounds are already available in the literature.

\subsection{Approximations and Bounds for the $Q$-Function}
A brief overview on the existing approximations and bounds for the Gaussian \mbox{$Q$-function} is presented herein with the focus on those with the exponential form. The 
approximations and bounds presented in \cite{Cody‎1969Jul,BorjessonSundberg1979Mar,Jelena2017,KaragiannidisLioumpas2007Aug,DyerDyer2008Apr,IsukapalliRao2008Sep,TellamburaAnnamalai2000Apr,ChenBeaulieu2009Feb,deAbreu2009Nov,deAbreu2012Sep,LopezBenitezCasadevall2011Apr,6139870,Jang2011Feb} have relatively complex mathematical forms and achieve high accuracy. 
Although some of them  may lead to closed-form expressions, which would be otherwise impossible to solve, e.g, the polynomial approximation in \cite{ChenBeaulieu2009Feb} succeeds in analytically evaluating the average symbol error rate of pulse amplitude modulation in log-normal channels,
the mathematical complexity of the aforementioned approximations make them still not quite convenient for algebraic manipulations in statistical performance analysis despite being accurate. For example, the approximation proposed by B{\"o}rjesson and Sundberg in \cite{BorjessonSundberg1979Mar} is very complicated and best suitable for programming purposes. Therefore, the simplest known family with the form of a sum of exponentials was proposed by Chiani {\em et al.}\cite{ChianiDardariSimon2003Jul}, to provide bounds and approximations based on the Craig's formula.

The expression for approximating or bounding $Q(x)$ by $\tilde{Q}(x)$ that is generally suitable for applications, where one needs to express average error probabilities for fading distributions with adequate accuracy, is written as \cite[Eq.~(8)]{ChianiDardariSimon2003Jul}
\begin{align}
\label{eq:3}
\tilde{Q}(x) \triangleq \sum_{n=1}^{N} a_n\exp\left(-b_n x^2\right)
\:\:\text{[for $x \geq 0$ only]}.
\end{align}
Chiani {\em et al.} use the monotonically increasing property of the integrand in~(\ref{eq:2}) and apply the rectangular integration rule to derive exponential upper bounds. Moreover, when using the trapezoidal rule with optimizing the center point to minimize the integral of relative error in an argument range of interest, an approximation with two exponential terms, $N=2$, is obtained.

Other exponential approximations and bounds are also available\cite{ChanCosmanMilstein2011Nov,LoskotBeaulieu2009Mar,OlabiyiAnnamalai2012Oct,7954666,8353222}. A coarse single-term exponential approximation is presented in\cite{ChanCosmanMilstein2011Nov} based on the Chernoff bound, and a sum of two or three exponentials is proposed in\cite{LoskotBeaulieu2009Mar}, which is known as the Prony approximation. Another approximation of the exponential form that shows good trade-off between computational efficiency and mathematical accuracy is proposed in\cite{OlabiyiAnnamalai2012Oct}. In~\cite{7954666}, the composite trapezoidal rule with optimally chosen number of sub-intervals is used. The authors in\cite{8353222} introduce a single-term exponential lower bound by using a tangent line to upper-bound the logarithmic function at some point which defines the tightness of the bound.

All of the aforementioned references propose approximations and bounds for the Gaussian \mbox{$Q$-function} and they can be also used as building blocks to approximate the powers or polynomials thereof. However, none of them directly derived approximations or bounds to evaluate the powers or polynomials of the \mbox{$Q$-function}, which arise frequently when analyzing various communication systems, e.g., error probability in quadrature amplitude modulation (QAM). 

\subsection{Applications of the Approximations and Bounds}

The above approximations and bounds have been implemented in the different areas of communication theory. We provide herein few examples from the literature. 
The approximations from\cite{IsukapalliRao2008Sep} and \cite{LopezBenitezCasadevall2011Apr} are used respectively to derive the frame error rate for a two-way decode-and-forward relay link in\cite{5601657}, and to analytically evaluate the average of integer powers of the $Q$-function over $\eta$--$\mu$ and $\kappa$--$\mu$ fading in\cite{7110787}. As for the exponential form, it is used in\cite{ChianiDardariSimon2003Jul} to compute error probabilities for space--time codes and phase-shift keying. Furthermore, (\ref{eq:3}) is used to derive the average bit-error rate for free-space optical systems in\cite{4786457} and the symbol error rate of phase-shift keying under Rician fading in\cite{McKayZanellaCollingsChiani2009Mar}.

In general, the elegance of the exponential approximation in (\ref{eq:3}) can be illustrated by
\begin{equation}
\resizebox{1\hsize}{!}{$\begin{aligned}
\int\,F\left(Q(f(\gamma))\right)\,Y(\gamma)\,\mathrm{d}\gamma
\approx\sum_{n}\,a_n\int\,\exp(-b_n [f(\gamma)]^2)\,Y(\gamma)\,\mathrm{d}\gamma,
\end{aligned}$}
\nonumber
\end{equation}
where $Y(\gamma)$ is some integrable function and $F\left(Q(f(\gamma))\right)$ is some well-behaved function of the \mbox{$Q$-function} that accepts a Taylor series expansion for $0 \leq Q(f(\gamma)) \leq \frac{1}{2}$. Above, the polynomial of $Q(f(\gamma))$ from the Taylor series of $F\left(Q(f(\gamma))\right)$ is approximated by (\ref{eq:3}), either directly or indirectly (by first approximating $Q(f(\gamma))$ by $\tilde{Q}(f(\gamma))$ and then expanding the polynomial of the sum), which results in the latter sum.

Evaluating the integral in the above summation is usually much easier than evaluating the integral in the original expression at the left-hand side. This idea is applied in~\cite{6777123}, when evaluating the average block error rate for Gamma--Gamma turbulence models under coherent binary phase-shift keying.
Taylor series can also be used to approximate $Y(\gamma)$ or parts of it~\cite{8425781,6777123}, eventually leading to closed-form expressions.
Finally, it is worth mentioning that increasing the number of exponential terms in the summation (\ref{eq:3}) will typically not increase the analytical complexity since summation and integration can be reordered in the expression under certain conditions and, hence, the integral is solved only once.

\subsection{Contributions and Organization of the Paper}

The objective of this paper is to develop new accurate approximations and bounds for the Gaussian $Q$-function and functions thereof. To that end, we adopt the exponential sum expression originally proposed in~\cite{ChianiDardariSimon2003Jul} and restated in~(\ref{eq:3}) and focus on the research problem of finding new, improved coefficients for it.\footnote{Throughout the paper, when referring to `our approximation/bound', we mean the existing sum expression~(\ref{eq:3}) from~\cite{ChianiDardariSimon2003Jul} with our new coefficients.} The coefficients developed herein will work as one-to-one replacements to those available in existing literature~\cite{ChianiDardariSimon2003Jul,ChanCosmanMilstein2011Nov,LoskotBeaulieu2009Mar,OlabiyiAnnamalai2012Oct,7954666,8353222}, but they offer significantly better accuracy and flexibility as well as generalization to various cases that have not been addressed before.

The major contributions of this paper are detailed as follows: 
\begin{itemize}

    \item We propose an original systematic methodology to optimize the set of coefficients $\{(a_n, b_n)\}_{n=1}^{N}$ of (\ref{eq:3}) to obtain increasingly accurate but tractable approximations for the \mbox{$Q$-function} with any $N$ in terms of the absolute or relative error, based on the minimax approximation theory, by which the global error is minimized when the corresponding error function is uniform.

    \item We further repurpose the methodology to find new exponential lower and upper bounds with very high accuracy that is comparable to, or even better than, the accuracy of other bounds of more complicated forms.
    
    \item We generalize our approximations and bounds to apply to polynomials and integer powers of the \mbox{$Q$-function}, or even implicitly to any generic function of the \mbox{$Q$-function} that accepts a Taylor series expansion.
    
    \item We show that the proposed minimax procedure reflects high flexibility in allowing for lower absolute or relative error at the expense of the other, or in allowing for higher accuracy in a specified range at the expense of less accuracy in the remaining ranges and a worse global error, by controlling weights assigned to the resulting non-uniform error function's extrema.

\end{itemize}
These contributions are verified by means of an extensive set of numerical results and an application example illustrating their accuracy and significance in communication theory.

The remainder of this paper is organized as follows. In Section \ref{sec:Preliminaries}, we present the mathematical preliminaries needed for the formulation of the research problem and proposed solutions. Section \ref{sec:Global Uniform} introduces our new approximations and bounds for the \mbox{$Q$-function}. 
Section \ref{sec:polynomial} presents our new approximations and bounds for the polynomials of the \mbox{$Q$-function}. The increasing accuracy of the novel solutions is demonstrated as well as comparisons with the best numerical alternatives and other known approximations having the same exponential form are presented in Section \ref{sec:results}. Concluding remarks are given in Section \ref{sec:Conclusion}.

\section{Preliminaries}
\label{sec:Preliminaries}

The case $x \geq 0$ is presumed throughout this article. The results can be usually extended to the negative real axis using the relation $Q(x) = 1 - Q(-x)$. Likewise, the following discussions focus solely on the Gaussian \mbox{$Q$-function} but the results directly apply also to the related error function $\operatorname{erf}(\cdot)$ and the complementary error function $\operatorname{erfc}(\cdot)$ through the identity $\operatorname{erfc}(x)= 1 - \operatorname{erf}(x) = 2\,Q\left(\sqrt{2}\,x\right)$, as well as to the cumulative distribution function $\Phi(\cdot)$ of a normal random variable  with mean~$\mu$ and standard deviation~$\sigma$ through the identity $\Phi(x)= 1 - Q\left(\frac{x-\mu}{\sigma}\right)$, which can be extended to $x<\mu$ using the relation $\Phi(x) = 1 - \Phi(2\mu-x) = Q\big(\frac{\mu-x}{\sigma}\big)$.

The approximations and bounds will be optimized shortly in terms of the absolute or relative error using the minimax approach, in which the possible error in the worst-case scenario (i.e., the maximum error over all $x$) is minimized. The baseline absolute and relative error functions\footnote{These should not be confused with the error function $\operatorname{erf}(\cdot)$.} are defined as
\begin{align}
\label{equ:6}
d(x) &\triangleq \tilde{Q}(x) - Q(x),\\
\label{equ:7}
r(x) &\triangleq \frac{d(x)}{Q(x)} = \frac{\tilde{Q}(x)}{Q(x)} - 1,
\end{align}
respectively, and the shorthand $e \in \{d,r\}$ represents both of them collectively in what follows. In particular, the tightness of some approximation or bound $\tilde{Q}(x)$ over the range $[x_0, x_{\infty}]$ is measured as
\begin{align}
\label{7}
\emax \triangleq \max_{x_0 \leq x \leq x_\infty} |e(x)|,
\end{align}
and the approximations and bounds for minimax error optimization are solved as
\begin{align}
\label{8}
\{(a_n^*, b_n^*)\}_{n=1}^{N} \triangleq \underset{\{(a_n, b_n)\}_{n=1}^{N}}{\operatorname{arg\,min}} \emax,
\end{align}
where $e(x) \geq 0$ for upper bounds and $e(x) \leq 0$ for lower bounds when $x \geq 0$. 

Our optimization method depends on the extrema of the error function (cf.\  Fig.~1), which occur at points $x_k$ where $e'(x_k)=0$, for which the derivatives are given by
\begin{align}
d'(x) &= \tilde{Q}'(x) - {Q}'(x),\\
r'(x) &= \frac{\tilde{Q}'(x)\,Q(x) -  \tilde{Q}(x)\,{Q}'(x)}{[Q(x)]^2}.
\end{align}
The derivatives of the approximation/bound in (\ref{eq:3}) and of the \mbox{$Q$-function} in (\ref{eq:1}) are
\begin{align}
\label{equ:11}
\tilde{Q}'(x) &= -2 \cdot \sum_{n=1}^{N} a_n\, b_n \, x \exp\left(-b_n x^2\right),\\
\label{equ:12}
{Q}'(x) &= -\frac{1}{\sqrt{2\pi}} \exp\left(-{\textstyle \frac{1}{2}} x^2\right),
\end{align}
respectively. Let us also note that the absolute error converges to zero when $x$ tends to infinity, i.e., $\lim\limits_{x\to\infty} d(x)=0$, whereas for the relative error, we have
\begin{align}
   {\lim\limits_{x\to\infty}{r(x)}=
    \begin{cases}
    \infty, &\text{when } \min\{b_n\}_{n=1}^N=\frac{1}{2},\\
   -1, &\text{otherwise.}
    \end{cases}}
    \label{13}
\end{align}
This renders some specific restrictions for all upper bounds and optimization w.r.t.\ the relative error as is shortly observed.

For reference, the Craig's formula in (\ref{eq:2}) can also be approximated
using various numerical integration techniques~\cite{numericalbook2}. This results in low-accuracy approximations or bounds of the same form as (\ref{eq:3}) with numerical coefficients that can be directly calculated from the weights and nodes of the corresponding numerical method.

\begin{figure}
\begin{center}
\includegraphics[width=.48\textwidth]{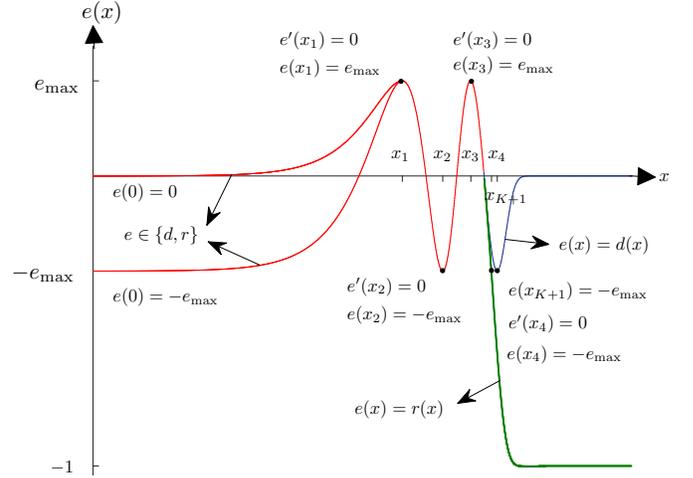}
\caption{The optimized minimax error function starts either from $e(0)=0$ or from $e(0)=-\emax$ and oscillates between local maximum and minimum values of equal magnitude; when considering relative error, this is possible only in a finite range of $x$ as opposed to global bounds obtained w.r.t.\ absolute error. The minimax criterion implies uniform error function with $w_k=1$.}
\end{center}
\label{fig:1}
\end{figure}

\section{Minimax Approximations and Bounds\\for the Gaussian $Q$-Function}
\label{sec:Global Uniform}

We adopt the weighted sum of exponential functions in (\ref{eq:3}) to express global minimax approximations and bounds for the Gaussian \mbox{$Q$-function}. In particular, according to Kammler in\cite[Theorem 1]{doi:10.1137/0713062}, the best approximation in which the maximum value of the corresponding error function is minimized to reach its minimax error, occurs when the error function is uniformly oscillating between maximum and minimum values of equal magnitude, as illustrated in Fig.~1. 

The original idea in our work is that one can describe the minimax error function by a set of equations, where the number of equations is equal to the number of unknowns. These equations describe the error function at the extrema points in which all of them have the same value of error and the derivative of the error function at these points is equal to zero. Our ultimate goal is then to find the optimized set of coefficients, $\{(a_n^*, b_n^*)\}_{n=1}^{N}$, that solves the formulated set of equations. In general, for problem formulation of $e\in\{d,r\}$,
\begin{align}
\label{equ:13}
	\begin{cases}
	e'(x_k) = 0,&\text{for } k=1,2,3,\ldots,K,\\
	\phantom{e'}\llap{$e$}(x_k) = (-1)^{k+1}\,w_k\,\emax,&\text{for } k=1,2,3,\ldots,K,
	\end{cases}
\end{align}
where $w_k$ is a potential weight for error at $x_k$ (set $w_k=1$ as default for uniform approximations/bounds) and $K$ is the number of the error function's extrema excluding the endpoints. Table~\ref{table1} summarizes the values of $K$ in terms of the number of sum terms $N$ for the different cases considered next.

In this study, we aim to minimize the global error over the whole positive $x$-axis, which is possible in terms of the absolute error. However, the relative error does not converge to zero when $x$ tends to infinity as seen in (\ref{13}). Thus, we must choose a finite interval on the $x$-axis, in which its right boundary, $x_{\infty}$, is equal to $x_{K+1}$ as will be discussed later. 
On the other hand, the left boundary of the $x$-range, $x_0$, is equal to zero for both error measures. In addition to $w_k,\, k=1,2,...,K$, the weight set also includes $w_0$ which occurs at $x_0$, and $w_{K+1}$ which occurs at $x_{K+1}$ for the relative error.

Although the minimum global absolute or relative error is obtained when the error function is uniform, the weight set that can be controlled is added throughout this article when formulating the approximation or bound problem to facilitate a compromise between $\dmax$ and $\rmax$ when tailoring it specifically for some application. The weight set can be even controlled to obtain better accuracy in some specified range of the argument. It should be mentioned that, in these cases, at least one of the weights has to be equal to one, representing the maximum error, and the remaining should be smaller and positive. When all of the weights are equal to one, the approximations and bounds are called uniform and they achieve the global minimax error as discussed earlier.

Two variations of equations can be formulated depending on whether the error starts from $e(0)=0$ or $e(0)=-w_0\,\emax$ as seen in Fig.~1. The importance of the former case comes from the fact that such approximation or upper bound gives the exact same value as the \mbox{$Q$-function} at $x=0$, resulting in a continuous function when extending it to the negative values of $x$. The latter case gives slightly better accuracy at the expense of the discontinuity that occurs at $x=0$.

\begin{table}[t]
\centering
\caption{Number of error extrema excluding endpoints needed to formulate the problem in terms of absolute or relative error.}
\begin{tabular}{|l|l|l|}
\hline
Error measure, $e$ & Type &Number of extrema\\
\hline
{\multirow{3}{*}{Absolute error, $d$}}
&Upper bound& $K=2N-1$\\
\cline{2-3}
&Approximation& $K=2N$\\
\cline{2-3}
&Lower bound&$K=2N$\\
\cline{2-3}
\hline
{\multirow{3}{*}{Relative error, $r$}}
& Upper bound& $K=2N-2$\\
\cline{2-3}
&Approximation&$K=2N-1$\\
\cline{2-3}
& Lower bound&$K=2N-1$\\
\cline{2-3}
\hline
\end{tabular}
\label{table1}
\end{table}

\subsection{Problem Formulation in Terms of Absolute Error}

Here we describe the formulation of the approximations and bounds of the \mbox{$Q$-function} when minimizing the global absolute error according to (\ref{7}) and (\ref{8}). The corresponding set of coefficients, $\{(a_n,b_n)\}_{n=1}^{N}$, in (\ref{eq:3}) are optimized as follows:
\begin{align}
\label{opt_d}
\{(a_n^*, b_n^*)\}_{n=1}^{N} \triangleq \underset{\{(a_n, b_n)\}_{n=1}^{N}}{\operatorname{arg\,min}} \max_{x \geq 0} \left|\tilde{Q}(x) - Q(x)\right|.
\end{align}

\subsubsection{Approximations}

The approximation's maximum absolute error is globally minimized when all local error extrema are equal to the global error extrema. The extrema occur where the derivative of the absolute error function is zero. For the produced error, all positive and negative extrema have the same value of error, i.e., $\dmax$. Moreover, we optimize (\ref{equ:6}) at $x_0=0$ for two variations: $d(0)=0$ or $d(0)=-w_0\,\dmax$, where $Q(0) = \frac{1}{2}$ and $\tilde{Q}(0) = \sum_{n=1}^{N} a_n$.

Therefore, we can formulate the approximation problem as
\begin{align}
\label{equ:14}
	\begin{cases}
	d'(x_k) = 0,\quad\quad\quad\quad\,\,\,\quad\quad\quad\text{for } k=1,2,3,\ldots,K,\\
	d(x_k) = (-1)^{k+1}\,w_k\,\dmax,\quad\,\text{for } k=1,2,3,\ldots,K,\\
	\begin{cases}
	\sum_{n=1}^{N} a_n=\frac{1}{2},&\text{when }d(0)=0,\\
		\sum_{n=1}^{N} a_n=\frac{1}{2}-w_0\,\dmax,&\text{when }d(0)=-w_0\,\dmax.
			\end{cases}
	\end{cases}
\end{align}
Although only the set $\{(a_n^*, b_n^*)\}_{n=1}^{N}$ is needed to construct the minimax absolute error function 
indicated by $e\in\{d,r\}$ together with $e(x)=d(x)$ in Fig.~1, other unknowns will also appear when solving the optimization problem in (\ref{opt_d}), which are $\{x_k\}_{k=1}^{K}$ and $\dmax$ for the uniform approximations and bounds.

The number of equations throughout this paper is always equal to the number of unknowns. For the minimax approximation in terms of absolute error, a set of $4N+1$ equations is constructed to solve $4N+1$ unknowns using $2N$ extrema points according to Table~\ref{table1}. Each extremum yields two equations; one expresses its value, and the other expresses the derivative of the error function at that point. An additional equation originates from evaluating the error function at $x_0$. This corresponds to either $e(0)=0$ or $e(0)=-\emax$ as indicated in Fig.~1. For any $N$, a solution to the system of equations yields $\{(a_n^*, b_n^*)\}_{n=1}^{N}$ that defines the minimax approximation, and we prove by construction that it exists. 

\subsubsection{Bounds}
For the bounds, we use the same approach as for the approximations with ensuring that $d(x)\leq0$ and $d(x)\geq0$ for the lower and upper bounds, respectively, when $x \geq 0$. The former results in $4N+1$ equations, with the optimized absolute error function starting from $d(0)=-w_0\,\dmax$, the maxima equal to zero and the minima equal to $-w_k\,\dmax$. On the other hand, the latter results in $4N$ equations with the corresponding error function starting from $d(0)=0$, the maxima equal to $w_k\,\dmax$ and the minima equal to zero, with forcing the lowest value in the set $\{b_n\}_{n=1}^N$ to be $\frac{1}{2}$, so that both error measures are always positive. Otherwise $r(x)$ will converge to a negative value as shown in (\ref{13}), $d(x)$ would be negative for large $x$ too, and we could not find an upper bound of the \mbox{$Q$-function}. Moreover, the derivative of the corresponding error function is equal to zero at all the $K$ extrema points for both types of bounds.

\subsection{Problem Formulation in Terms of Relative Error}

Here we describe the formulation of the exponential approximations and bounds of the \mbox{$Q$-function} when minimizing the global relative error defined by (\ref{equ:7}). We optimize the corresponding set of coefficients, $\{(a_n,b_n)\}_{n=1}^{N}$, as follows:
\begin{align}
\label{opt_r}
\{(a_n^*, b_n^*)\}_{n=1}^{N} \triangleq \underset{\{(a_n, b_n)\}_{n=1}^{N}}{\operatorname{arg\,min}}\max_{0 \leq x \leq x_{K+1}} \left|\frac{\tilde{Q}(x)}{Q(x)} - 1\right|.
\end{align}
Unlike the absolute error, the relative error does not converge to zero when $x$ tends to infinity as shown in (\ref{13}). This is why we must limit the minimax approximation in terms of the relative error to the finite range by choosing $x_{\infty}=x_{K+1}$, as opposed to $x_{\infty}\to\infty$ in the case of absolute error. This yields
\begin{align}
\begin{cases}
\label{relative_endpoint}
	r\left(x_{K+1}\right) = \phantom{-}w_{K+1}\,\rmax,&\text{for upper bounds,}\\
	r\left(x_{K+1}\right) = - w_{K+1}\,\rmax,&\text{otherwise.}
\end{cases}
\end{align}
Hence, the relative error function is minimized globally over $[0,x_{K+1}]$.
This can be seen by the case where $e(x)=r(x)$ in Fig.~1, in which the point $x_{K+1}$ is chosen so that its corresponding error value is equal to $-\rmax$.

\subsubsection{Approximations}

In regard to the relative error, the same approach as for the absolute error is implemented herein in order to construct the minimax approximations with the corresponding uniform error function illustrated by $e\in\{d,r\}$ together with $e(x)=r(x)$ in Fig.~1. 
A set of $4N$ equations originates from the $2N-1$ extrema and the two endpoints, which are $x_0$ and $x_{K+1}$. It is noted that, $r'(x_{K+1})\neq0$ and only one equation can be acquired from this point, since the minimax approximation herein is limited to the range $0 \leq x \leq x_{K+1}$. Therefore, the optimized coefficients for the two variations are found by solving the following set of equations:
\begin{align}
\label{19}
	\begin{cases}  
	r'(x_k) = 0,\quad\quad\quad\,\,\,\,\,\,\,\,\quad\quad\quad\quad\text{for } k=1,2,3,\ldots,K,\\
r(x_k) = (-1)^{k+1}\,w_k\,\rmax,\quad\,\,\,\,\,\,\text{for } k=1,2,3,\ldots,K,\\
			\begin{cases}
		\sum_{n=1}^{N} a_n=\frac{1}{2},&\text{when } r(0)=0,\\
	\sum_{n=1}^{N} a_n=\frac{1}{2}-\frac{1}{2}\,w_0\,\rmax,&\text{when } r(0)=-w_{0}\rmax,\\
			\end{cases}\\
				r(x_{K+1}) =          - w_{K+1}\,\rmax.
		\end{cases}
\end{align}

\subsubsection{Bounds} 

We optimize the lower and upper bounds for $0 \leq x \leq x_{K+1}$ in terms of the relative error using the same problem formulation as for the absolute bounds but with $4N$ equations in case of lower bounds, and $4N-1$ equations in case of upper bounds, and by substituting $d$ by $r$, in addition to enforcing (\ref{relative_endpoint}) that describes the error function at $x_{k+1}$.

\subsection{Proof by Construction: Solutions for $N=1,2,3,\ldots,25$}

We prove the existence of the proposed solutions to (\ref{opt_d}) and (\ref{opt_r}) by construction, i.e., numerically solving (\ref{equ:14}) and (\ref{relative_endpoint}), (\ref{19}). In particular, we implemented the set of equations of each of the considered variations in Matlab and used the {\tt fsolve} command with equal number of equations and unknowns to find the optimized set of coefficients $\{(a^*_n, b^*_n)\}_{n=1}^{N}$, where the main challenge was to choose heuristic initial guesses. For the initial guesses of lower values of $N$, we used iteratively random values for $\emax$, $\{(a_n, b_n)\}_{n=1}^{N}$ and $\{x_k\}_{k=1}^{K}$ with $K$ as given in Table~\ref{table1}, along the process of finding their optimal values that solve the proposed research problem. After reaching certain $N$ which is enough to form a relation between the previous values, we constructed a pattern to predict their successive values for higher values of $N$.
 
The sets of optimized coefficients are solved herein up to $N=25$ for the novel minimax approximations and bounds as well as released to public domain in a supplementary digital file with  $x_{K+1}$ ranging from $1$ to $10$ in steps of $0.1$ for the relative error. Nevertheless, let us illustrate the sets of optimized coefficients of the absolute error for $d(0)=-\dmax$ and $N=2,3,4$ in Table \ref{table4}, in addition to the set of optimized coefficients of the relative error in the case where $r(0)=0$, $x_{K+1}=6$ and $N=20$, for quick reference.

Our optimized coefficients yield very accurate approximations that outperform all the existing ones in terms of the global error. For example, for $N=2$, our approximation yields $\dmax= 9.546\cdot 10^{-3}$ and the reference approximations~\cite{ChianiDardariSimon2003Jul}, \cite{LoskotBeaulieu2009Mar} and \cite{OlabiyiAnnamalai2012Oct}, yield $\dmax= 1.667\cdot 10^{-1}, 1.450\cdot 10^{-1}$ and $1.297\cdot 10^{-1}$, respectively. The accuracy can be increased even further by increasing $N$. For example, the tabulated coefficients of the relative error for $N=20$ render a tight uniform approximation in terms of the relative error while satisfying $\tilde{Q}(0)=Q(0)=\frac{1}{2}$. Namely, $|r(x)| \leq \rmax^* < 2.831\cdot 10^{-6}$ when $x \leq 6$ and $|r(x_k)| = \rmax^*$ at all the $K=39$ local maximum error points. This approximation is also tight in terms of the absolute error since $|d(x)| \leq \dmax < 1.416\cdot 10^{-6}$ for all $x \geq 0$ and the largest local error maxima are observed when $x \ll 1$ while $|d(x)| \ll \dmax$ for $x > 1$.

\begin{table}[h]
\centering
\caption{The set of optimized coefficients of the absolute error for $d(0)=-\dmax$ and $N=2,3,4$, and the set of optimized coefficients of the relative error for $r(0)=0$, $x_{K+1}=6$ and $N=20$.}
\begin{tabular}{|l|l|l|l|}
 \hline
$N$&$n$&$a_n^*$ & $ b_n^*$\\
\hline
\hline
$2$&$1$  &$3.736889599671366\mathrm{e-}1$& $8.179084584179674\mathrm{e-}1$ \\\cline{2-4}
 & $2$   &  $1.167651897698837\mathrm{e-}1$   &$1.645047046852372\mathrm{e+}1$     \\\cline{2-4}
\hline
\hline
$3$& $1$&$3.259195350781647 \mathrm{e-}1$&
$7.051797307608448\mathrm{e-}1$\\\cline{2-4}
 &  $2$   & $1.302528627687561\mathrm{e-}1$    &     $5.489376068647640\mathrm{e+}0$  \\\cline{2-4}
  & $3$    & $4.047435009465072\mathrm{e-}2$  &      $1.335391071637174\mathrm{e+}2$ \\\cline{2-4}
\hline
\hline
$4$& $1$ & $2.936683276537767\mathrm{e-}1$& $6.517755981618476\mathrm{e-}1$\\\cline{2-4}
 & $2$ &  $1.357580421878250\mathrm{e-}1$   &       $3.250040490513459\mathrm{e+}0$\\\cline{2-4}
  &    $3$ &  $5.245255757691102\mathrm{e-}2$   &    $3.186882707224491\mathrm{e+}1$   \\\cline{2-4}
 &   $4$  & $1.673209873360605\mathrm{e-}2$    &     $7.786613983601425\mathrm{e+}2$   \\\cline{2-4}
\hline
\hline
$20$&$1$&$7.558818716991463\mathrm{e-}2$& $5.071654316592885\mathrm{e-}1$\\
 \cline{2-4}
&$2$&$7.283303478836754\mathrm{e-}2$&$5.678040654656637\mathrm{e-}1$\\
\cline{2-4}
&$3$&$6.886155063785772\mathrm{e-}2$&
$7.104625738749141\mathrm{e-}1$\\
 \cline{2-4}
&$4$&$6.439172935348138\mathrm{e-}2$& $9.994060383297402\mathrm{e-}1$\\
 \cline{2-4}
&$5$&$5.779242444673264\mathrm{e-}2$&$ 1.601184575755943\mathrm{e+}0$\\
\cline{2-4}
& $6$&$4.808415837769939\mathrm{e-}2$&$ 2.928772702717808\mathrm{e+}0$\\
\cline{2-4}
&$7$&$3.692309273438261\mathrm{e-}2$&$ 6.019071014437780\mathrm{e+}0$\\
 \cline{2-4}
&$8$&$2.656563850645104\mathrm{e-}2$&$ 1.358210951915055\mathrm{e+}1$\\
 \cline{2-4}
& $9$&$1.820530043799255\mathrm{e-}2$&$ 3.304520236491907\mathrm{e+}1$\\
  \cline{2-4}
& $10$&$1.201348364882034\mathrm{e-}2$&$ 8.584892772825742\mathrm{e+}1$\\
  \cline{2-4}
& $11$&$7.675500579336059\mathrm{e-}3$&$ 2.375751011169581\mathrm{e+}2$\\
  \cline{2-4}
 &$12$&$4.755522827095319\mathrm{e-}3$&$ 7.025476884457923\mathrm{e+}2$\\
  \cline{2-4}
&$13$&$2.853832378872099\mathrm{e-}3$&$ 2.237620299200472\mathrm{e+}3$\\
 \cline{2-4}
 &$14$&$1.652925274323080\mathrm{e-}3$&$ 7.776239381556935\mathrm{e+}3$\\
  \cline{2-4}
 &$15$&$9.183202474880042\mathrm{e-}4$&$ 3.007617539336614\mathrm{e+}4$\\
  \cline{2-4}
&$16$&$4.846308477760495\mathrm{e-}4$&$ 1.334789827558299\mathrm{e+}5$\\
 \cline{2-4}
& $17$&$2.391717111298367\mathrm{e-}4$&$ 7.146006517383908\mathrm{e+}5$\\
  \cline{2-4}
& $18$&$1.074573496224467\mathrm{e-}4$&$ 5.056149657406912\mathrm{e+}6$\\
  \cline{2-4}
&$19$&$4.174113678130675\mathrm{e-}5$&$ 5.790627530626244\mathrm{e+}7$\\
 \cline{2-4}
&$20$&$1.229754587599716\mathrm{e-}5$&$ 2.138950747557404\mathrm{e+}9$\\
 \hline
\end{tabular}
\label{table4}
\end{table}


\section{Approximations and Bounds for\\Polynomials of the $Q$-Function}
\label{sec:polynomial}

In this section, we generalize the novel minimax optimization method presented in Section \ref{sec:Global Uniform}, to derive approximations and bounds for any polynomial of the \mbox{$Q$-function} and any integer power of the \mbox{$Q$-function} as a special case. In fact, this method can be applied to expressing approximations and bounds for many well-behaved functions of the \mbox{$Q$-function} using Taylor series expansion, in which it is represented as an infinite sum of terms. Therefore, Taylor series is a polynomial of infinite degree\cite{taylor} that one needs to truncate to get a Taylor polynomial approximation of degree $P$.

In general, any $P$th degree polynomial of the \mbox{$Q$-function} is expressed as 
\begin{align}
\label{23}
\Omega\left(Q(x)\right)\triangleq\sum_{p=0}^P c_p\,Q^p(x),
\end{align}
where $\{c_p\}_{p=0}^{P}$ are constants and called the polynomial coefficients. 
In particular, the novel optimization methodology is extended to such polynomials by directly approximating/bounding $\Omega\left(Q(x)\right)$ by $\tilde{Q}_{\Omega}(x)$ that has the same exponential form as $\tilde{Q}(x)$ in (\ref{eq:3}). We optimize the coefficient set, $\{(a_n, b_n)\}_{n=1}^{N}$, in order to minimize the maximum absolute or relative error of the polynomial, which results in a uniform error function as described before.

The absolute and relative error functions for any polynomial of the \mbox{$Q$-function} are defined respectively as
\begin{align}
\label{abs}
d_{\Omega}(x) &\triangleq \tilde{ Q}_{\Omega}(x) - \sum_{p=0}^P c_p\,Q^p(x),\\
\label{rel}
r_{\Omega}(x) &\triangleq \frac{d_\Omega(x)}{\sum_{p=0}^P c_p\,Q^p(x)} = \frac{\tilde{Q}_{\Omega}(x)}{\sum_{p=0}^P c_p\,Q^p(x)} - 1.
\end{align}
The derivatives of the error functions are 
\begin{equation}
\label{abs_der}
d'_{\Omega}(x) = \tilde{Q}'_{\Omega}(x) -\sum_{p=1}^P p\,c_p Q^{p-1}Q'(x),
\end{equation}
\begin{equation}
\resizebox{.98\hsize}{!}{$\begin{aligned}
\label{rel_der}
r'_{\Omega}(x) =
\frac{\tilde{Q}'_{\Omega}(x)\sum_{p=0}^P c_p\,Q^p(x)-\tilde{Q}_{\Omega}(x)\sum_{p=1}^P p\,c_p Q^{p-1}Q'(x)}{\left[\sum_{p=0}^{P} c_p\,Q^p(x)\right]^2},
\end{aligned}$}
\end{equation}
where $\tilde{Q}'_{\Omega}(x)$ has the same expression as $\tilde{Q}'(x)$ in (\ref{equ:11}) and ${Q}'(x)$ is given by (\ref{equ:12}).

Following the procedure explained in Section~\ref{sec:Global Uniform}, and using the mentioned definitions, approximations/bounds for polynomials of the \mbox{$Q$-function} are formulated in terms both error measures. More specifically, what applies to error functions with the \mbox{$Q$-function} described by (\ref{opt_d})--(\ref{19}) also applies herein, with replacing $\sum_{n=1}^{N} a_n=\frac{1}{2}$ by $\sum_{n=1}^{N} a_n=\sum_{p=0}^P (\frac{1}{2})^p\,c_p$ for the absolute and relative errors of the approximations that start from  $e(0)=0$ and for the upper bounds. Furthermore, one should replace $\sum_{n=1}^{N} a_n=\frac{1}{2}-w_0\,\dmax$ by $\sum_{n=1}^{N} a_n=\sum_{p=0}^P (\frac{1}{2})^p\,c_p-w_0\,\dmax$ for the absolute error and $\sum_{n=1}^{N} a_n=\frac{1}{2}-\frac{1}{2}w_0\rmax$ by $\sum_{n=1}^{N} a_n=\sum_{p=0}^P (\frac{1}{2})^p\,c_p-\sum_{p=0}^P (\frac{1}{2})^p\,c_p\,w_0\,\rmax$ for the relative error of the approximations that start from $e(0)=-w_0\,\emax$ and for lower bounds.

\subsection{Special Case: Integer Powers of the $Q$-Function}
\label{sec:Resultspower}

In general, any polynomial of the \mbox{$Q$-function} as per (\ref{23}) is a linear combination of non-negative integer powers of the \mbox{$Q$-function}. The integer powers themselves are important special cases in communication theory, where they appear frequently on their own.
To that end, one may derive the optimized approximations and bounds for them by simply setting the coefficient $c_p$ of the required power $p$ in (\ref{23})--(\ref{rel_der}) to one and the remaining to zero while following exactly the same optimization procedure as explained above for the general case of polynomials. It should also be mentioned that, for the upper bounds, $\min\{b_n\}_{n=1}^N=\frac{p}{2}$. We refer to the approximations and bounds of this special case by $\tilde{Q}_p(\cdot)$ to differentiate it from the general case of polynomials.
 
In the coefficient data that we release to public domain along with this paper, the sets of optimized coefficients $\{(a_n^*,b_n^*)\}_{n=1}^{N}$ for the approximations/bounds of the exponential form shown in (\ref{eq:3}) are numerically solved with $p = 1$, $2$, $3$, $4$ and $N=1,2,\ldots,25$ for the novel minimax approximations and bounds with $x_{K+1}$ ranging from $1$ to $10$ in steps of $0.1$ for the relative error. However, the provided approximations and bounds can be extended to any value of $p$.

If not approximating directly, the approximations/bounds for any polynomial of the \mbox{$Q$-function} with $N$ terms can be obtained  by using the integer powers' approximations/bounds (including the first power) as follows:
\begin{align}
\Omega\left(Q(x)\right)&\approx\sum_{p=0}^P c_p\, \tilde{Q}_{p,N_p}(x)\nonumber\\
\label{24}
&=\sum_{p=0}^P c_p\,\prod_{l=1}^{L}\tilde{Q}_{p_l,N_{p_l}}(x)\nonumber\\
& =\sum_{p=0}^P c_p\,\sum_{n_{p_1}=1}^{N_{p_1}}\sum_{n_{p_2}=1}^{N_{p_2}}...\sum_{n_{p_L}=1}^{N_{p_L}} \prod_{l=1}^{L}a_{n_{p_l}}[l]\nonumber\\ &\times\exp{\left(-\left(\sum_{l=1}^{L}b_{n_{p_l}}[l]\right)\,x^2\right)},
\end{align}
where $\sum_{l=1}^L{p_l}=p$, $\prod_{l=1}^L N_{p_l}=N_p$, $\sum_{p=0}^{P} N_p=N$, and $a_{n_p}[l]$, $b_{n_p}[l]$ are the coefficients of $\tilde{Q}_{p_l,N_{p_l}}(x)$. The ultimate number of terms in (\ref{24}) may be less than $N$ if some of them can be combined. The above implies also that the approximations/bounds of any integer power of the \mbox{$Q$-function} with $N_p$ terms can be obtained using the product rule.

\subsection{Application Example: Evaluation of the Average SEP in Optimal Detection of $4$-QAM in Nakagami-$m$ Fading}
\label{sec:Resultssep}

Let us emphasize on the elegance of (\ref{eq:3}) for approximating or bounding the \mbox{$Q$-function}, its integer powers or any polynomial thereof by giving an application example of average error probabilities over fading channels. In general, they are obtained for coherent detection in most cases by evaluating
\begin{equation}
\bar{P}_E=\int_0^{\infty} \Omega\left(Q(\alpha\,\sqrt{\gamma})\right)\,\psi_{\gamma}(\gamma)\mathrm{d}\gamma,
\label{eq:application3}
\end{equation}
where $\Omega\left(Q(\alpha\,\sqrt{\gamma})\right)$ is some polynomial of the \mbox{$Q$-function} as per (\ref{23}) and refers to the error probability conditioned on the instantaneous signal-to-noise ratio (SNR), i.e., $\gamma$, with $\psi_{\gamma}(\gamma)$ being its probability density function, and $\alpha$ is a constant that depends on the digital modulation and detection techniques. Substituting our approximation into the above equation yields
\begin{subequations}
\begin{align}
\label{eq:application4.a}
\bar{P}_E&\approx\sum_{n=1}^N a_n \int_0^{\infty} \exp(-b_n\,\alpha^2\,\gamma)\,\psi_{\gamma}(\gamma)\mathrm{d}\gamma\\
&=\sum_{n=1}^N\,a_n\Theta_\gamma(-b_n\alpha^2),
\label{eq:application4.b}
\end{align}
\end{subequations}
where $\Theta_\gamma(s)=\int_{0}^{\infty}\exp(s\gamma)\,\psi_{\gamma}(\gamma)\,\mathrm{d}\gamma$ is the moment generating function associated with the random variable $\gamma$.  

Let us next evaluate the average symbol error probability (SEP) in optimal detection of \mbox{$4$-QAM} over Nakagami-$m$ fading channels, under which it is often hard to derive closed-form expressions for error probabilities if $m$ is not an integer.
Thus, we first solve exponential approximations and bounds for the conditional SEP in \mbox{$4$-QAM} that is a second-order polynomial of the \mbox{$Q$-function} as follows \cite[Eq.~8.20]{digibook}:
\begin{align}
\label{33}
P_E\left(\gamma\right) &=2\,Q\left({\sqrt{\gamma}}\right)-Q^{2}\left(\sqrt{\gamma}\right).
\end{align}
By comparing to (\ref{23}), $c_0=0,\,c_1=2$, and $c_2=-1$. This SEP is approximated by $\tilde{Q}_{\Omega}(x)$ as described above.
Finally, we substitute the gamma probability distribution in (\ref{eq:application4.a}) and evaluate the integral using \cite[Eq.~3.351.3]{tableofseries} as
\begin{align}
\bar{P}_E&=
\frac{m^m}{\overline{\gamma}^m\Gamma(m)}\sum_{n=1}^{N}a_n\int_0^{\infty} \gamma^{m-1}\exp\left(-\gamma\left(b_n+\frac{m}{\overline{\gamma}}\right)\right)\mathrm{d}\gamma\nonumber\\
&=\frac{m^m}{\overline{\gamma}^m}\sum_{n=1}^{N}a_n\,\left(b_n+\frac{m}{\overline{\gamma}}\right)^{-m},
\end{align}
where $m$ defines the fading parameter, ranging from $0.5$ to $\infty$, $\overline{\gamma}$ is the average SNR, and $\Gamma(\cdot)$ denotes the gamma function.

The sets of optimized coefficients $\{(a_n^*,b_n^*)\}_{n=1}^{N}$ for the approximations and bounds of the conditional SEP in \mbox{$4$-QAM} were solved for $N=1,2,\ldots,25$ for the minimax approach in terms of both error measures. Table \ref{table5} shows an example of the coefficients optimized in terms of the absolute error in the case where $d_{\Omega}(0)=-\dmax$ and $N=5$. These render a tight uniform approximation with $|d_{\Omega}(x)| \leq \dmax^* < 6.84\cdot 10^{-4}$.

\begin{table}[t]
\centering
\caption{The set of optimized coefficients of the absolute error for $d_{\Omega}(0)=-\dmax$ and $N=5$.}
\begin{tabular}{|l|l|l|}
 \hline
$n$&$a_n^*$ & $ b_n^*$\\
 \hline
 $1$&$4.920547396876422\mathrm{e-}1$& $5.982476003750250\mathrm{e-}1$\\
 \hline
$2$&$1.587491012166297\mathrm{e-}1$&$2.024383866054074\mathrm{e+}0$\\
 \hline
$3$&$6.460001610510117\mathrm{e-}2$&
$1.323465438792062\mathrm{e+}1$\\
 \hline
$4$&$2.567521272080907\mathrm{e-}2$& $1.314581690889673\mathrm{e+}2$\\
 \hline
$5$&$8.236936034796302\mathrm{e-}3$&$ 3.211202445024321\mathrm{e+}3$\\
\hline
\end{tabular}
\label{table5}
\end{table}

The computational and/or analytical complexity using our approximations and bounds for the integer powers and the polynomials of the \mbox{$Q$-function} is much less than using any other approximation from the literature, in which none of them has proposed approximations or lower/upper bounds for the powers or the polynomials of the \mbox{$Q$-function}. Therefore, directly substituting the SEP polynomial by our exponential approximations is more tractable than evaluating it by applying reference approximations to (\ref{33}).

\section{Numerical Results and Discussion}
\label{sec:results}

Let us next compare the proposed approximations and bounds with the existing ones having the same exponential form, in addition to the best approximations among the  different numerical integration techniques. The optimized sets of coefficients, $\{(a_n^*, b_n^*)\}_{n=1}^{25}$ for the cases considered in this paper, all in terms of both absolute and relative error, are constructed in this paper to form round $37\,000$ coefficient sets in total. Due to Matlab's fixed (64-bit) floating-point precision, some other programming software with adjustable precision is required to pursue the proposed minimax approach for finding approximations and bounds for values of $N$ much beyond $25$. This is because some $a_n$ become very small when the corresponding $b_n$ become very large resulting in underflow when computing $a_n\exp\left(-b_n x^2\right)$ numerically for (\ref{eq:3}).

To begin, we plot the minimax absolute error versus minimax relative error for $p=1, 2, 3, 4$, and $N=1, 2, 3, ... ,25$ of the approximation starting from $e(0)=0$, in Fig.~2, with  showing $x_{K+1}$ ranging from $1$ to $10$ for $N=5, 10, 15, 25$ in terms of relative error. The other types of approximations and the lower/upper bounds follow similar behaviour as the one shown in Fig.~2. It is clear from the figure that, as the number of exponential terms increases, the minimax absolute and relative error decrease significantly.

\begin{figure}
\begin{center}
\includegraphics[width=0.48\textwidth]{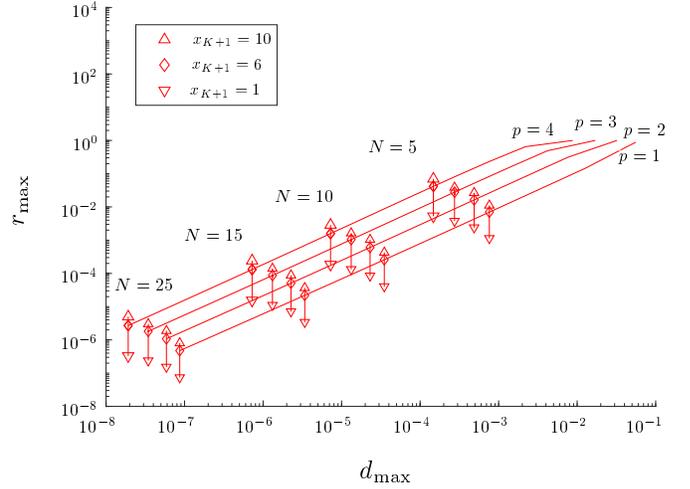}
\label{fig:10}
\caption{Optimal absolute error versus optimal relative error for the first four powers of the \mbox{$Q$-function} for the approximations starting from $e(0)=0$. The two-sided vertical arrows indicate $\rmax$ for $x_{K+1}$ ranging from $1$ to $10$.}
\end{center}
\end{figure}

\begin{figure}[t]
\centering
{\includegraphics[width=0.48\textwidth]{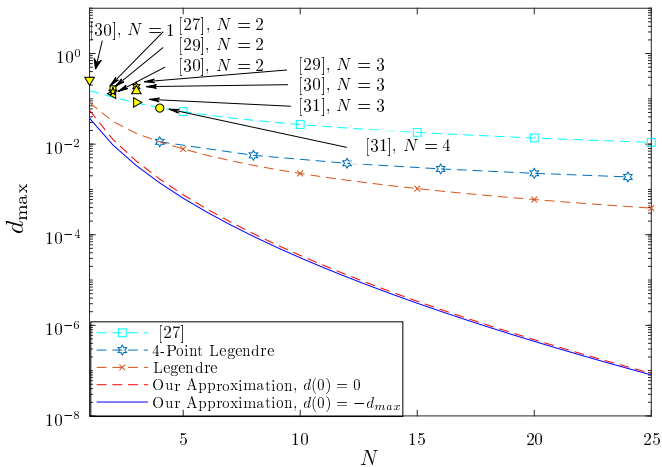}}
\caption{{Comparison of the absolute error between our approximations and those obtained using \cite{ChianiDardariSimon2003Jul,OlabiyiAnnamalai2012Oct,LoskotBeaulieu2009Mar} and\cite{7954666}, as well as those calculated using Legendre rule and its $4$-point composite version.}}
\end{figure}

For reference, we have investigated the different numerical integration techniques and their $h$-point composites (up to $h=4$) that can be implemented to approximate the Gaussian $Q$-function as a weighted sum of exponentials in terms of both absolute and relative errors. However, we only include the Legendre rule and its four-point composite formula in Fig.~3, where they achieve the least global error among all the other numerical methods and their composites, respectively, along with the two types of the proposed minimax approximations. In addition, the global error values of the existing approximations of the same form are also calculated and plotted in the same figure for specific number of terms, namely, $N=1, 2, 3, 4$, where $\tilde{Q}(\cdot)$ is expressed using one exponential in \cite{OlabiyiAnnamalai2012Oct}, two exponentials in~\cite{ChianiDardariSimon2003Jul}, \cite{LoskotBeaulieu2009Mar} and \cite{OlabiyiAnnamalai2012Oct}, three exponentials in~\cite{LoskotBeaulieu2009Mar},\cite{OlabiyiAnnamalai2012Oct} and~\cite{7954666}, and four exponentials in~\cite{7954666}. 
The composite right-rectangular rule, which was used to approximate $Q(\cdot)$ in~\cite{ChianiDardariSimon2003Jul}, is also plotted for comparison. In Fig.~3, we only include the absolute error since the relative error illustrates similar results, and only the maximum error over $x\geq 0$ is compared.

It is evident from the figure that our approximations outperform all of the existing approximations as well as those obtained from numerical integration in terms of the global error, and as the number of terms increases, even better accuracy is obtained. In contrast, we can see that the numerical methods are converging slowly, causing the number of terms required by the numerical integration to be much higher than that required by our approximations in order to achieve the same level of error.

Table \ref{table3} compares the values of $N$ between the proposed approximations and the best integration rules that achieve certain absolute error levels. Clearly, our approximations are much more tractable than any other numerical approximation in terms of the global error, where only a few exponential terms are needed to achieve high accuracy. For the non-composite Legendre rule, when applied to approximate the \mbox{$Q$-function}, the error will start to oscillate for $N>41$ and eventually converge to infinity. This implies that Legendre approximations are not reliable and cannot achieve high level of accuracy. After illustrating the efficiency of our proposed approximations in terms of the global error, we further verify the accuracy for the whole considered range of the positive argument by comparing the relative error function obtained when applying our approximations and the existing ones for $N=2$ and $N=4$ as shown in Fig.~4.
In addition to the fact that our approximations have the least global error, their accuracy surpass all the reference approximations over the range $[0, 0.4]$ and attain comparable accuracy for $x>0.4$.

For the ranges, where other approximations have better accuracy, the error function can be reshaped in such a way that the accuracy over the specified range is improved at the cost of less accuracy in the other ranges and, hence, increased global error. We do that by controlling the weights of the error function's extrema of our approximations when setting the problem conditions.

As an example, let us consider the problem conditions in (\ref{19}) that formulates the relative error shown in Fig.~4. We can increase the accuracy of the approximation which has three extrema for $N=2$ and starts from $r(0)=-w_0\,\rmax$ over the range $[-2, 14]$, by controlling the weights of the extrema to be $w_0=1,w_1=w_2=w_3=w_4=1/10$, $w_4$ is the weight at the right boundary of the interval of optimization. This example is illustrated in the figure by the solid-diamond line. We can see that the error has decreased to be more accurate in the specified range and outperforms the other reference approximations over most of the range. However, the global error has increased substantially. This demonstrates how our approximations' and bounds' accuracy can be tailored for specific ranges of values, depending on their application.

The accuracy of our upper and lower bounds was investigated in terms of both error measures but only the relative error is shown in Fig.~5 to save space. It is obvious that our bounds not only have the least global error but they also outperform the other exponential bounds presented in \cite{ChianiDardariSimon2003Jul, ChanCosmanMilstein2011Nov}. Moreover, over a wide range of the argument, our bounds have even better accuracy than the other bounds of more complicated forms. For instance, our lower bound is the best over the whole positive range $x>0$. On the other hand, our upper bound has better accuracy than that of\cite{deAbreu2009Nov} and comparable accuracy to\cite{Jang2011Feb}, although \cite{deAbreu2012Sep} is more accurate over the range $[0, 3.5]$, where it has a more complex form.

\begin{table}[t]
\centering
\caption{Comparison between $N$ values for the proposed approximations and both composite and non-composite Legendre integration rules that achieve certain absolute error level.}
\begin{tabular}{|c|c|c|c|}
\hline
\shortstack{Absolute\\error} &  \shortstack{$N$ for approx.\ \\ with $d(0)=0$}  &\shortstack{$N$ for composite \\ Legendre rule}& \shortstack{$N$ for non-composite \\ Legendre rule}\\
\hline
$1\cdot10^{-2}$ & $2$ & $4$  & $4$\\
\hline
$1\cdot10^{-3}$ & $4$ & $44$ & $15$ \\
\hline
$1\cdot10^{-4}$ & $8$ & $452$ & $41$ \\
\hline
$1\cdot10^{-5}$ & $12$ & $3504$ & $-$ \\
\hline
\end{tabular}
\label{table3}
\end{table}

As mentioned earlier, we can achieve better absolute or relative error at the expense of the other by controlling the weights of the extrema. We test the trade-off behaviour herein by starting from the uniform relative error with equal weights and gradually decreasing the weights' values, $w_k$, for $k=1, 2,...,K$ while maintaining $w_{K+1}=1$. The maximum obtained relative and absolute error values are measured and plotted in Fig.~6 for $N=1, 2,...,10$. The cross marker in the figure refers to the minimax error obtained when formulating the minimax approximation in terms of absolute error, for any $N$. In the same way, the plus marker refers to the minimax error obtained when formulating the minimax approximation in terms of relative error, for any $N$. We can see from Fig.~6 that as the absolute error decreases, the relative error increases, forming smooth transition and a trade-off between the two error measures. Other transition lines can be formed between the extremes based on how the weight set is controlled.

\begin{figure}
  \centering
 {\includegraphics[width=0.47\textwidth]{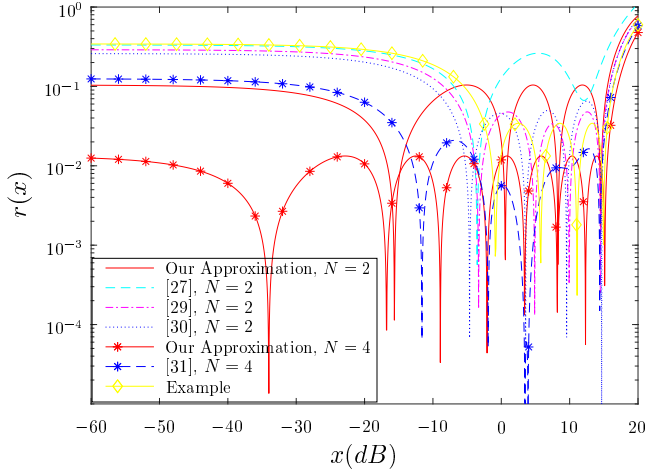}}
  \caption{Comparison among our approximations and the references approximations,  \cite{ChianiDardariSimon2003Jul,OlabiyiAnnamalai2012Oct,LoskotBeaulieu2009Mar} and\cite{7954666} for $N=2$ and $N=4$ in terms of the relative error. }
\end{figure}

\begin{figure}
\begin{center}
\label{fig:bounds}
\includegraphics[width=0.48\textwidth]{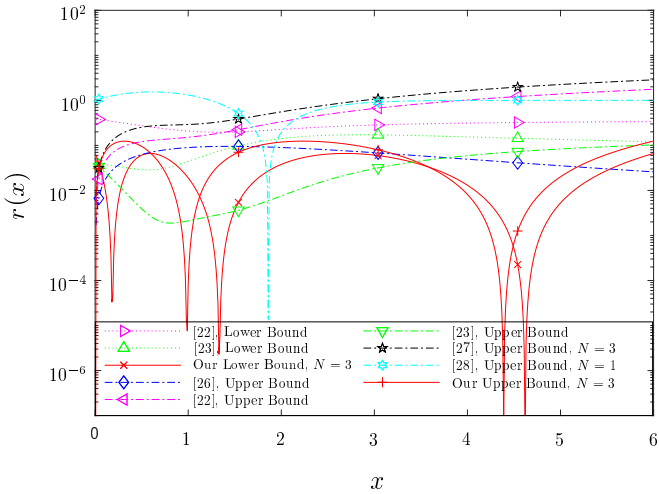}
\caption{Comparison between our bounds and the references bounds\cite{ChianiDardariSimon2003Jul,ChanCosmanMilstein2011Nov,deAbreu2009Nov,deAbreu2012Sep,Jang2011Feb} in terms of relative error.}
\end{center}
\end{figure}

For the relative error, the effect of changing the value of $x_{K+1}$ is illustrated in Fig.~7. As the value of $x_{K+1}$ increases, $r_{max}$ increases too for the approximations and bounds, achieving worse accuracy. Furthermore, like noted before, higher values of $N$ result in highly improved accuracy as can be seen in the figure, in which the relative error for $N=25$ is several orders of magnitude lower than for $N=5$.

 \begin{figure}
\begin{center}
\label{fig:trade-off}
\includegraphics[width=0.47\textwidth]{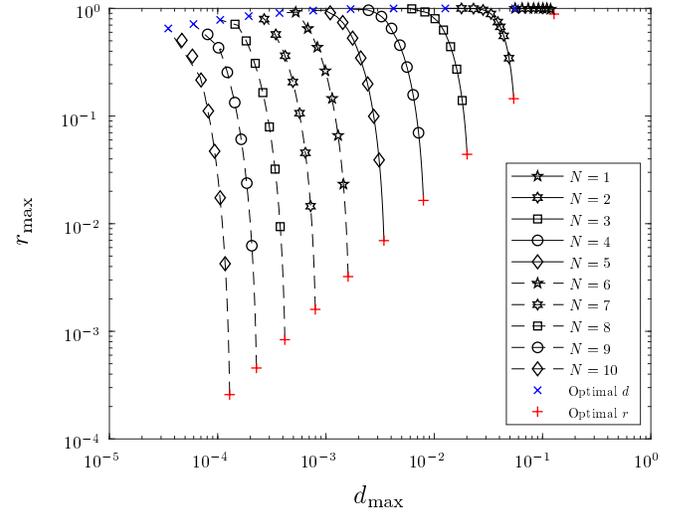}
\caption{The trade-off between the absolute and relative error. To obtain better absolute error than obtained when optimizing the relative error, the weight set when formulating the optimization problem can be controlled to achieve less absolute error but with increased relative error, and vice versa.}
\end{center}
\end{figure}

 \begin{figure}
\begin{center}
\includegraphics[width=0.47\textwidth]{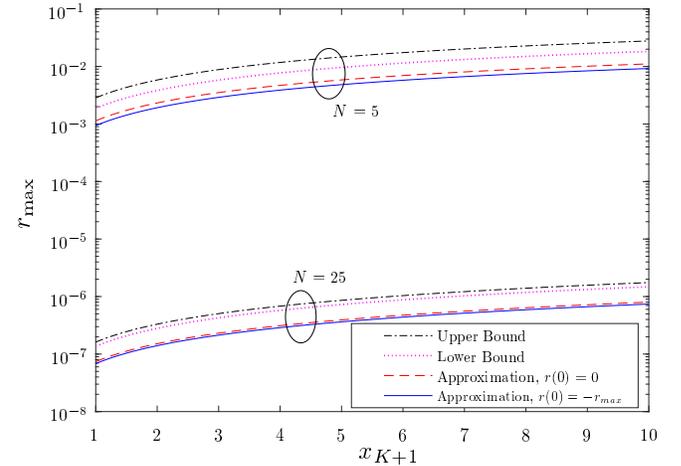}
\caption{The effect of changing $x_{K+1}$ on the relative error of the proposed approximations and bounds, for $N=5$ and $N=25$.}
\end{center}
\label{fig:5}
\end{figure}

In Fig.~8, we compare the absolute error of the proposed approximations and bounds for the third power of the \mbox{$Q$-function}, with the error calculated using (\ref{24}) for all $N$. The minimum error among all errors obtained using all the possible combinations of $N_{p_l}$, $l=1,\ldots,L$ is considered in this comparison for each combination set of the \mbox{$Q$-function} whose powers add to three.
It is noted that representing the integer powers of the \mbox{$Q$-function} as weighted sum of exponentials using (\ref{eq:3}), is more accurate and simpler than representing it using the different combinations. 

Finally, approximating SEP in (\ref{33}) directly using (\ref{eq:3}) in the coherent detection of $4$-QAM is compared in Fig.~9(a) with those obtained using the different combinations when applying (\ref{24}). The direct solutions give increasingly higher absolute and relative accuracy as expected.  Figure~9(b) together with Table~\ref{table:qam} compares the accuracy of the corresponding average SEP in $4$-QAM when evaluated for different values of the fading parameter $m$ using our exponential approximation, $\tilde{Q}_{\Omega}(\cdot)$, with the optimized coefficients that are listed in Table~\ref{table5}, and the other reference exponential approximations. The results demonstrate excellent agreement over the entire range of average SNR between the exact average SEP and our approximation that is very tight even for lower values of SNR, in contrast to the references that are accurate only at higher SNRs. Furthermore, the tightness of our approximation is preserved when changing the value of $m$, while the approximation from \cite{ChianiDardariSimon2003Jul} is accurate only for small values of $m$. It should be noted that, when we substitute the reference approximations with two terms in (\ref{33}), we get a five-term exponential approximation for the SEP. As the number of exponential terms increases, our approximation becomes virtually exact, outperforming all the existing approximations as seen in Table \ref{table:qam} with already $N=10$.

\begin{figure}
\begin{center}
\includegraphics[width=0.48\textwidth]{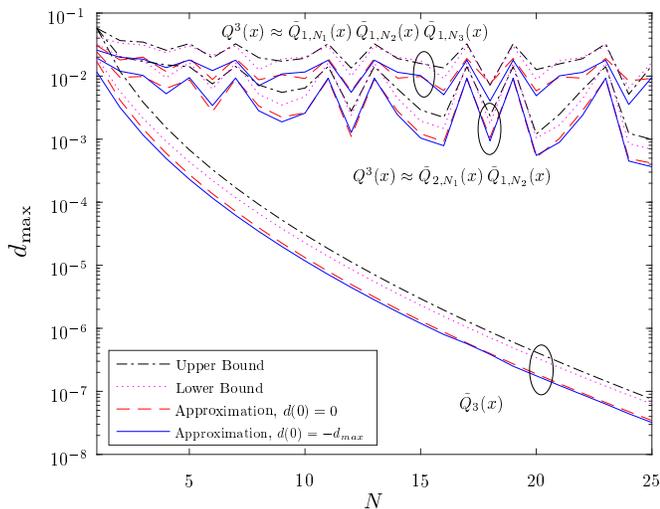}
\caption{Comparison of the absolute error of the proposed exponential approximations and bounds for $p=3$, with the minimum error among all errors obtained using all the possible combinations as given in (\ref{24}).}
\end{center}
\label{fig:6}
\end{figure}

 \begin{figure}
\begin{center}
 \subfigure[]{\includegraphics[width=0.48\textwidth]{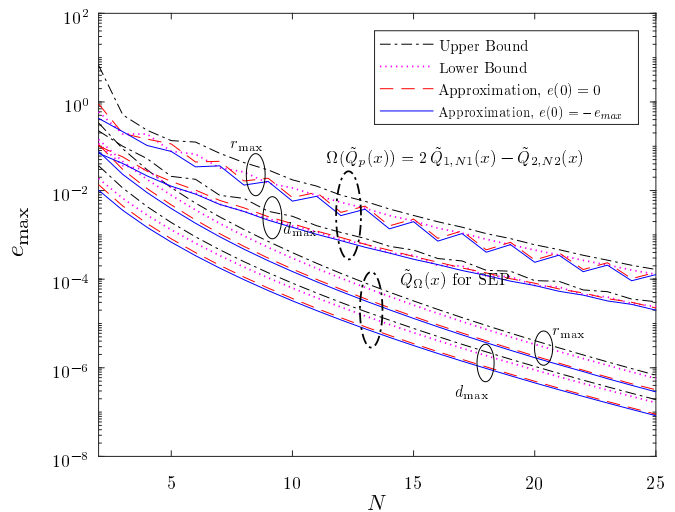}}
 \subfigure[]{\includegraphics[width=0.46\textwidth]{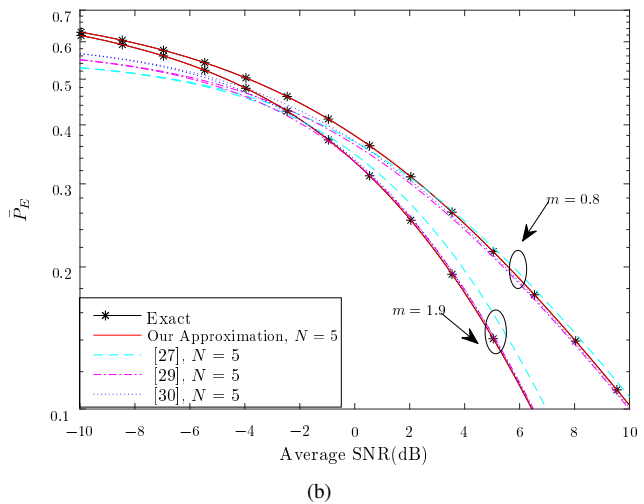}}
\caption{(a) Comparison of both absolute and relative error of the proposed exponential approximations and bounds for SEP in $4$-QAM, with those obtained by applying (\ref{24}). (b) Average SEP plots for $4$-QAM over Nakagami-$m$ using our approximation and the reference exponential approximations for $N = 5$.}
\end{center}
\label{fig:8}
\end{figure}

\begin{table}[t]
\centering
\caption{Comparison of accuracy of average SEP for $4$-QAM over Nakagami-$m$ fading.}
\begin{tabular}{l|l|l|l|l}
 \hline
\multicolumn{5}{c}{For $m= 0.8$}\\
  \hline
Exact & $0.530436$ & $0.379629$ & $0.216681$& $0.101863$ \\
\hline
\cite{ChianiDardariSimon2003Jul}, $N=5$ & $0.478317$ & $0.368463$ & $0.220893$& $0.106127$ \\
\hline
\cite{LoskotBeaulieu2009Mar}, $N=5$ & $0.487660$ & $0.363137$ & $0.211014$& $0.099769$ \\
\hline
\cite{OlabiyiAnnamalai2012Oct}, $N=5$ & $0.499954$ & $0.369262$ & $0.213148$ & $0.100468$\\
\hline
\shortstack{\scalebox{.9}[1.0]{Our approx., $N=5$}}& $0.530440$ & $0.379629$ & $0.216629$ & $0.101753$\\
\hline
\shortstack{\scalebox{.9}[1.0]{Our approx., $N=10$}}& $0.530436$ & $0.379629$ & $0.216680$ & $0.101859$\\
\hline
\hline
\multicolumn{5}{c}{For $m= 1.9$}\\
\hline
Exact & $0.509397$ & $0.333819$ & $0.142200$& $0.034658$ \\
\hline
\cite{ChianiDardariSimon2003Jul}, $N=5$ & $0.474948$ & $0.346634$ & $0.160587$& $0.040565$ \\
\hline
\cite{LoskotBeaulieu2009Mar}, $N=5$ & $0.482288$ & $0.333805$ & $0.144435$& $0.035039$ \\
\hline
\cite{OlabiyiAnnamalai2012Oct}, $N=5$ & $0.493865$ & $0.337280$ & $0.143837$ & $0.034678$\\
\hline
\shortstack{\scalebox{.9}[1.0]{Our approx., $N=5$}}& $0.509432$ & $0.333780$ & $0.142188$ & $0.034474$\\
\hline
\shortstack{\scalebox{.9}[1.0]{Our approx., $N=10$}}& $0.509398$ & $0.333819$ & $0.142200$ & $0.034652$\\
\hline
\hline
$\overline{\gamma}$ (in dB)&$-5$&$0$ & $5$ & $10$\\
\hline
\end{tabular}
\label{table:qam}
\end{table}

\section{Conclusions}
\label{sec:Conclusion}

This paper proposed accurate and tractable approximations, lower bounds and upper bounds for the Gaussian \mbox{$Q$-function} and any polynomial of the \mbox{$Q$-function} as a weighted sum of exponential functions. The novel sets of coefficients of the sum terms are optimally solved in minimax sense to minimize the global absolute or relative error of approximations/bounds, where in the limit of a larger number of terms, they approach very close to their corresponding exact functions. Moreover, we show that the weights set to the extrema of the error function can be controlled to compromise between the absolute and the relative error. The significantly (i.e., by several orders of magnitude) improved accuracy of the proposed expressions with optimized coefficients has been demonstrated by comparing the results with approximations from numerical integration and other existing approaches.

\bibliographystyle{IEEEtran}
\bibliography{refs}

\begin{IEEEbiography}[{\includegraphics[width=1in,height=1.25in,clip,keepaspectratio]{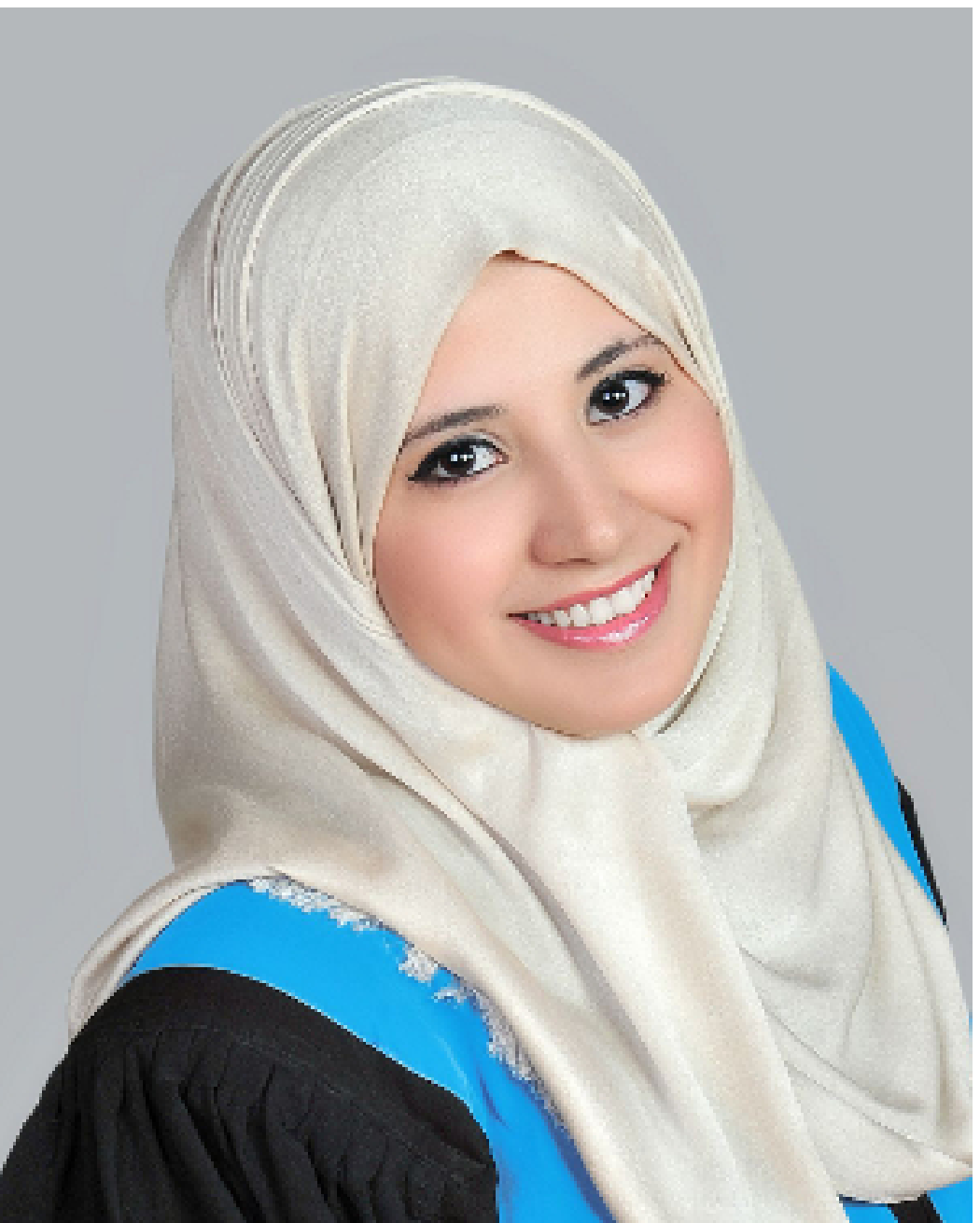}}]{Islam M. Tanash}
received the B.Sc.\ and M.Sc.\ degrees in electrical engineering from Jordan University of Science and Technology (JUST), Irbid, Jordan in 2014 and 2016, respectively. She is currently a PhD student and doctoral researcher at the Faculty of Information Technology and Communication Sciences, Tampere University, Finland. Her research interests include the areas of communications theory, wireless networks, and wireless systems security.
\end{IEEEbiography}

\begin{IEEEbiography}[{\includegraphics[width=1in,height=1.25in,clip,keepaspectratio]{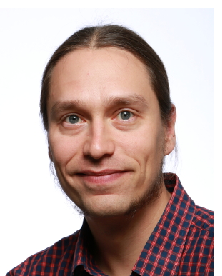}}]{Taneli Riihonen}(S'06--M'14)
received the D.Sc.\ degree in electrical engineering (with distinction) from Aalto University, Helsinki, Finland, in August 2014. He is currently an Assistant Professor (tenure track) at the Faculty of Information Technology and Communication Sciences, Tampere University, Finland. He held various research positions at Aalto University School of Electrical Engineering from September 2005 through December 2017. He was a Visiting Associate Research Scientist and an Adjunct Assistant Professor at Columbia University in the City of New York, USA, from November 2014 through December 2015. He has been nominated eleven times as an Exemplary/Top Reviewer of various IEEE journals and is serving as an Editor for \textsc{IEEE Wireless Communications Letters} since May 2017. He has previously served as an Editor for \textsc{IEEE Communications Letters} from October 2014 through January 2019. He received the Finnish technical sector's award for the best doctoral dissertation of the year and the EURASIP Best PhD Thesis Award 2017. His research activity is focused on physical-layer OFDM(A), multiantenna, relaying and full-duplex wireless techniques with current interest in the evolution of beyond 5G systems.
\end{IEEEbiography}

\vfill

\end{document}